\documentclass[
aps,prd,
reprint,
superscriptaddress,% author affiliations via superscript numbers
nofootinbib,
amsmath,amssymb,
floatfix
]{revtex4-2}
\usepackage[utf8]{inputenc}
\usepackage{upgreek, stix, physics, dsfont, bm, mathrsfs} % fonts and symbols

\usepackage{graphicx, xcolor, tikz, tikzscale}
\definecolor{maroon}{HTML}{800020}
\definecolor{darkcyan}{HTML}{008060}
\definecolor{theme1}{rgb}{0.9, 0.36, 0.054} % orange-red
\definecolor{theme2}{rgb}{0.3652, 0.4278, 0.7583} % blue-purple
\definecolor{theme3}{rgb}{0.9451, 0.5939, 0} % yellow-orange
\definecolor{theme4}{rgb}{0.6460, 0.2532, 0.6851} % purple
\definecolor{theme5}{rgb}{0.2858, 0.56, 0.4508} % aqua green
\definecolor{theme6}{rgb}{0.7, 0.336, 0} % brown
\definecolor{theme7}{rgb}{0.4915, 0.3451, 0.8} % cool purple
\definecolor{theme8}{rgb}{0.7179, 0.5687, 0} % dull yellow

\graphicspath{{figures/}}
\usepackage[caption=false]{subfig}

\usepackage[colorlinks]{hyperref} % hyperlinking
\hypersetup{colorlinks=true, allcolors=maroon}
\usepackage[capitalise]{cleveref} % avoid \ref, use \cref instead
\crefname{section}{Sec.}{Secs.} % to match PRD crossref style
\crefname{equation}{}{} % remove 'eq." from equations

\newcommand{\mat}[1]{\mathsf{#1}}

\begin{document}

\title{Scale limited fields and the Casimir effect}

\author{\v{S}imon~Vedl}
\email[Please direct correspondence to: ]{simon.vedl@hdr.mq.edu.au}
\affiliation{Department of Physics and Astronomy,
Macquarie University, Sydney, NSW 2109, Australia}
\affiliation{Sydney Quantum Academy, Sydney, NSW 2000, Australia}
\affiliation{ARC Centre of Excellence in Engineered Quantum Systems,
Macquarie University, Sydney, NSW 2109, Australia}

\author{Daniel~J.~George}
\affiliation{Department of Physics and Astronomy,
Macquarie University, Sydney, NSW 2109, Australia}
\affiliation{Sydney Quantum Academy, Sydney, NSW 2000, Australia}
\affiliation{ARC Centre of Excellence in Engineered Quantum Systems,
Macquarie University, Sydney, NSW 2109, Australia}

\author{Gavin~K.~Brennen}
\affiliation{Department of Physics and Astronomy,
Macquarie University, Sydney, NSW 2109, Australia}
\affiliation{ARC Centre of Excellence in Engineered Quantum Systems,
Macquarie University, Sydney, NSW 2109, Australia}

\begin{abstract}
     We revisit the calculation of the Casimir effect from the perspective of scale limited resolutions of quantum fields.
     We use the continuous wavelet transform to introduce a scale degree of freedom and then restrict it to simulate either an observational or fundamental limitation of resolution.
     The Casimir force is derived in this setting for a free complex massless scalar field between two infinite plates with both Dirichlet and periodic boundary conditions.
     The dependence of the force on the choice of wavelet and size of scale cutoff is extensively discussed for several examples of wavelets. 
\end{abstract}

\maketitle

%--------------------------------------------------%
\section{Introduction}
The Casimir effect~\cite{casimir1948a} is perhaps the most direct illustration of  how the zero-point energy of a free quantum field theory (QFT) can have observable consequences in the presence of boundary conditions.
Like many calculations in QFT, it is plagued by divergences, and a variety of regularisation techniques are used to obtain the accepted finite result~\cite{Hagen2001}.
Often a frequency cutoff, or bandlimit, is introduced.
This can be motivated by the physical characteristics of the boundary walls, such as the Debye cutoff of the frequency response of the boundary to the electromagnetic field. It can also arise from more fundamental reasons, such as the incompleteness of QFT at length scales on the order of the Planck length, or potentially larger-scale limits such as the minimum length scales described by generalised uncertainty principles (GUPs)~\cite{kempf1995}.

While it is the objective of every good regularisation technique to provide the same result in the continuum limit, we would also like to have an understanding of the nature of corrections arising from either fundamental limits on length scales, or due to constraints imposed by the physical construction of observables within the QFT.
The preferred tool for analysis of scale-dependent phenomena is wavelet theory.
Originally developed for the study of seismic signals~\cite{goupillaud1984,grossmann1984}, it has found broad application in physics and mathematics.
Within QFT, the discrete wavelet transform has found use in the study of Green's functions and renormalization flow  ~\cite{battle1999,altaisky2016,altaisky2018}, where the wavelet scale variable provides a natural coordinate to represent coarse grained information. Indeed, the Daubechies wavelet scale functions are themselves the solution to a renormalization group equation.
Other applications for wavelet-based analysis in QFT include the study of entanglement via a multi-scale wavelet representation~\cite{george2022}, the holographic principle~\cite{singh2016, brennen2015}, and tensor networks, namely, the Multi-scale Entanglement Renormalization Ansatz (MERA)~\cite{PhysRevLett.116.140403,Witteveen2022}, which can be used to describe ground states of quantum systems at criticality.

Wavelet regularization was first used for calculation of the Casimir effect by Altaisky and Kaputkina~\cite{altaisky2011} (see also~\cite{altaisky2013}).
Using first-order Hermitian continuous wavelets, they found a lowest-order correction that was attractive after expanding in the ratio of scale cutoff to plate separation.
Similar to momenta-based regularization as in bandlimited QFT~\cite{pye2015}, wavelet regularization restricts the range of scales on which physical processes---such as pair creation and the exchange of virtual particles---can occur, which results in a QFT free from divergences.

In this paper, we expand this wavelet regularization program, and find the lower-order corrections due to the scale cutoff depend on the wavelet family used, and are not always attractive.
This demonstrates care must be taken when inferring observable consequences of scale cutoffs, and may provide guidance to probing emergent effects of scale cutoffs in other contexts within QFT.
We begin in~\cref{sec:WaveletTransform} with a brief introduction to the continuous wavelet transform, providing the necessary mathematical foundation for the applications to follow.
The Casimir effect is then derived in~\cref{sec:CasimirEffect} using wavelet regularization.
Finally, in~\cref{sec:Detection} the form of the Casimir force is discussed for different choices of wavelets.
The objective is to explore the dependence of the resulting force on the choice of wavelet family.

%--------------------------------------------------%

%--------------------------------------------------%
\section{Continuous wavelet transform}\label{sec:WaveletTransform}

The continuous wavelet transform of an $L^2$ function $\phi(x)$ in $d$ dimensions under the \emph{wavelet} (\emph{aperture}) \emph{function} $w(x)$ is defined by the formula
\begin{equation}
    \phi_{a,\theta}(x) = (w_{a,\theta,x},\phi)=\int_{\mathbb{R}^d} \dd[d]{x'} w^*_{a,\theta,x}(x')\phi(x^{\prime}),
\end{equation}
where $w_{a,\theta,x}(x')$ is the unitary action of the similitude group $\mathbb{SIM}(d)$ on the wavelet function $w(x)$
\begin{equation}
    w_{a,\theta,x}(x') = \frac{1}{a^{d/2}}w\left(\mat{R}^{-1}(\theta)\frac{x'-x}{a}\right).
\end{equation}
The similitude group $\mathbb{SIM}(d)$ of the $d$-dimensional Euclidean space preserves the Euclidean norm up to multiplication by a constant.
It can be parameterised by a $d$-dimensional vector $x$ representing translations, an element $\theta$ of the group $SO(d)$ corresponding to rotations, and by a positive real number $a$ for scaling.
Here $(\bullet,\bullet)$ is the usual inner product of $L^2$ and $\mat{R}$ the natural representation of $SO(d)$ on $\mathbb{R}^d$.
The wavelet function $w$ must satisfy the admissibility condition
\begin{align}
    C_w &= \frac{1}{\norm{w}^2}\int_{\mathbb{SIM}(d)}\dd{\mu(a,\theta,x)} \abs{(w,w_{a,\theta,x})}^2 \\
    &= \int_{\mathbb{R}^d}\frac{\abs{\Tilde{w}(k)}^2}{\abs{k}^d}\dd[d]{k}<\infty, \label{eq:AdmissibilityCondition}
\end{align}
where $\Tilde{w}$ is the Fourier transform of $w$ and $ \dd{\mu(a,\theta,x)}$ is the left-invariant Haar measure on $\mathbb{SIM}(d)$:
\begin{gather}
    \dd{\mu(a,\theta,x)} = \frac{\dd{a}\dd{\mu(\theta)}\dd[d]{x}}{a^{d+1}}.
\end{gather}
Interestingly, wavelets can be viewed as generalised coherent states associated with this group \cite{antoine2015}.
If the admissibility condition is satisfied, the wavelet transform can be inverted using the reconstruction formula
\begin{equation}
    \phi(x) = \frac{1}{C_w}\int_{\mathbb{SIM}(d)}\dd{\mu(a,\theta,x')}w_{a,\theta,x'}(x)\phi_{a,\theta}(x').
\end{equation}
The wavelet transform is an isometry, which means that it preserves the inner product
\begin{equation}
    \int_{\mathbb{R}^d}\dd[d]{x}\; \phi^*(x)\psi(x) = \frac{1}{C_w}\int_{\mathbb{SIM}(d)}\dd{\mu(a,\theta,x)}\phi^*_{a,\theta}(x)\psi_{a,\theta}(x).
\end{equation}
Restricting integration over the scale from $(0,+\infty)$ to $(A,+\infty)$, or equivalently, projecting to the subspace of scale-limited functions, modifies the original inner product to 
\begin{equation}
    (\phi,\psi)_A = \int_{\mathbb{R}^{2d}}\frac{\dd[d]{x}\,\dd[d]{x'}}{A^d}\phi^*(x)f\left(\frac{\abs{x-x'}}{A}\right)\psi(x'),
\end{equation}
where the function $f$ is called the \emph{cutoff function}.
Its interpretation is more clear in the Fourier image
\begin{equation}
    (\phi,\psi)_A = \int_{\mathbb{R}^d}\frac{\dd[d]{k}}{(2\pi)^d} \Tilde{\phi}^*(k)\Tilde{f}(A\abs{k})\Tilde{\psi}(k),
\end{equation}
where
\begin{equation}
    \Tilde{f}(k) = \frac{1}{C_w}\int_{\abs{k'}>k}\frac{\abs{\Tilde{w}(k')}^2}{\abs{k'}^d}\dd[d]{k'}
    \label{eq:CutoffFunction}
\end{equation}
is essentially the admissibility condition integrated for all momenta, except for a ball of radius $k$ centred at the origin.
The effect of $\Tilde{f}$ is the attenuation of high momenta, hence the name cutoff function.
%--------------------------------------------------%

%--------------------------------------------------%
\section{The Casimir effect}\label{sec:CasimirEffect}

The Casimir effect \cite{casimir1948,casimir1948a} is observable when two conductive plates are very close to each other.
There is an attractive force between them proportional to $1/s^4$, where $s$ is the separation of the plates.
The force can be explained by the difference in the vacuum energy of the electromagnetic field between the plates and the vacuum energy outside.
In the following, a simple model of a free complex massless scalar field is considered.
Such a field satisfies the wave equation (henceforth we set $c=\hbar\equiv 1$):
\begin{equation}
    \Box\phi = \frac{\partial^2\phi}{\partial t^2}-\Delta\phi = 0.
\end{equation}
Without specifying the boundary conditions, the solution to this equation can be written as a superposition of plane waves
\begin{equation}
    \phi(t,\bm{x}) = 
    \int \frac{\dd[3]{\bm{k}}}{(2\pi)^3} \frac{1}{2\omega_k}
    \left(a(\bm{k})e^{-i(\omega_k t-\bm{k}\cdot \bm{x})} + b^*(\bm{k})e^{i(\omega_k t-\bm{k}\cdot \bm{x})} \right),
\end{equation}
where $\omega_k = \sqrt{\bm{k}^2} = \abs{\bm{k}}$.
After quantisation, the coefficient $a(\bm{k})$ is promoted to an annihilation operator $\hat{a}(\bm{k})$ of the particle, and $b^*(\bm{k})$ is promoted to a creation operator $\hat{b}^\dag(\bm{k})$ of the antiparticle.

The objective is then to calculate the vacuum energy which is given by the vacuum expectation value of the Hamiltonian
\begin{equation}
    \hat{H} = \int \dd[3]{\bm{x}}\left( \hat{\pi}^\dag(t,\bm{x})\hat{\pi}(t,\bm{x}) + \nabla\hat{\phi}^\dag(t,\bm{x})\cdot\nabla\hat{\phi}(t,\bm{x})\right),
\end{equation}
where the conjugate fields are given by
\begin{align}
    \hat{\pi}^\dag(t,\bm{x}) &= \dot{\hat{\phi}}(t,\bm{x})\equiv \partial_t\hat{\phi}(t,\bm{x}),\\
    \hat{\pi}(t,\bm{x}) &= \dot{\hat{\phi}}^\dag(t,\bm{x})\equiv \partial_t \hat{\phi}^\dag(t,\bm{x}),
\end{align}
and the field and conjugate field operators satisfy the canonical equal-time commutation relations:
\begin{align}
    [\hat{\phi}(t,\bm{x}),\hat{\pi}(t,\bm{x}')] = i\delta(\bm{x}-\bm{x}'),\\
    [\hat{\phi}^\dag(t,\bm{x}),\hat{\pi}^\dag(t,\bm{x}')] = i\delta(\bm{x}-\bm{x}'),
\end{align}
with the other commutators vanishing.
The field operators have dimensions of inverse length, and the conjugate field operators dimensions of inverse length squared.
The wavelet transform of $\hat{\phi}$ is performed in the spatial coordinates
\begin{equation}
    \hat{\phi}_{a,\theta}(t,\bm{x}) = \int_{\mathbb{R}^3} \dd[3]{\bm{x}'} \frac{1}{a^{3/2}}w^*\left(\mat{R}^{-1}(\theta)\frac{\bm{x}^{\prime}-\bm{x}}{a}\right)\hat{\phi}(t,\bm{x}^{\prime}),
\end{equation}
and the other fields are transformed so that the Hermitian conjugation is respected.
For example,
\begin{equation}
    \hat{\phi}^\dag_{a,\theta}(t,\bm{x}) = \int_{\mathbb{R}^3} \dd[3]{\bm{x}'} \frac{1}{a^{3/2}}w\left(\mat{R}^{-1}(\theta)\frac{\bm{x}^{\prime}-\bm{x}}{a}\right)\hat{\phi}^\dag(t,\bm{x}^{\prime}).
\end{equation}
The scale-dependent field in the wavelet picture also admits a mode expansion
\begin{multline}
    \hat{\phi}_{a,\theta}(t,\bm{x})
    = \int \frac{\dd[3]{\bm{k}}}{(2\pi)^3} \frac{1}{2\omega_k}
    \Big(\hat{a}_{a,\theta} (\bm{k})e^{-i(\omega_k t-\bm{k}\cdot \bm{x})}
    \\ + \hat{b}_{a,\theta}^\dag (\bm{k})e^{i(\omega_k t-\bm{k}\cdot \bm{x})}\Big),
\end{multline}
where the scaled creation and annihilation operators are essentially the wavelet transform of the original creation and annihilation operators in momentum space:
\begin{align}
    \hat{a}_{a,\theta}(\bm{k}) 
    &= a^{3/2} \, \Tilde{w}^*(-a\mat{R}^{-1}(\theta)\bm{k}) \, \hat{a}(\bm{k}),
    \\
    \hat{b}_{a,\theta}(\bm{k}) 
    &= a^{3/2} \, \Tilde{w}(a\mat{R}^{-1}(\theta)\bm{k}) \, \hat{b}(\bm{k}),
\end{align}
satisfying modified commutation relations
\begin{align}
    [\hat{a}_{a,\theta}(\bm{k}),\hat{a}^\dag_{a',\theta'}(\bm{k}')]
    & =(2\pi)^3 2\omega_k(aa')^{3/2} \, \delta(\bm{k}-\bm{k}')
    \\ \nonumber
    & \quad \times \Tilde{w}^*(-a\mat{R}^{-1}(\theta)\bm{k}) \, \Tilde{w}(-a' \mat{R}^{-1}(\theta')\bm{k}'),
    \\
    [\hat{b}_{a,\theta}(\bm{k}),\hat{b}^\dag_{a',\theta'}(\bm{k}')]
    &=(2\pi)^3 2\omega_k (aa')^{3/2} \, \delta(\bm{k}-\bm{k}')
    \\ \nonumber
    & \quad \times \Tilde{w}(a\mat{R}^{-1}(\theta)\bm{k}) \, \Tilde{w}^*(a' \mat{R}^{-1}(\theta')\bm{k}'). 
\end{align}
Therefore, as in the canonical representation, the Hamiltonian in the wavelet representation can be written in terms of creation and annihilation operators
\begin{multline}
    \hat{H} = \frac{1}{2}
    \int \frac{\dd{a}\dd{\mu(\theta)}\dd[3]{\bm{k}}}{C_w a^4(2\pi)^3}\omega_{k}
    \Big(\hat{a}_{a,\theta}^\dag(\bm{k})\hat{a}_{a,\theta}(\bm{k})
    + \hat{a}_{a,\theta}(\bm{k})\hat{a}_{a,\theta}^\dag(\bm{k}) 
    \\
    + \hat{b}_{a,\theta}^\dag(\bm{k})\hat{b}_{a,\theta}(\bm{k}) 
    + \hat{b}_{a,\theta}(\bm{k})\hat{b}_{a,\theta}^\dag(\bm{k})\Big).
    \label{eq:WaveletHamiltonian}
\end{multline}

\subsection{Periodic boundary conditions}
The presence of the conducting plates is modelled by imposing periodic boundary conditions on the field
\begin{equation}
    \hat{\phi}(t,x,y,0) = \hat{\phi}(t,x,y,s)
\end{equation}
which restricts the $z$ component of the momentum to values $k_z = \frac{2\pi n}{s}$, where $n$ is an integer.
The Hamiltonian is then \cref{eq:WaveletHamiltonian} with the replacements
\begin{equation}
    k_z \to \frac{2\pi n}{s},\quad \int\frac{\dd[3]{\bm{k}}}{(2\pi)^3}\to \sum_{n\in\mathbb{Z}}\int\frac{\dd[2]{k_\parallel}}{(2\pi)^2}.
\end{equation}
The energy density inside the plates is defined as the vacuum expectation value of the Hamiltonian $\rho_0(s)=\frac{1}{s}\ev{\hat{H}}{0}$.
We introduce a scale cutoff by assuming that the vacuum cannot be excited into modes below a certain length scale $A$:
\begin{equation}
    \hat{a}^\dag_{a,\theta}(\bm{k})\ket{0} = 0,\quad \hat{b}^\dag_{a,\theta}(\bm{k})\ket{0} = 0, \quad a < A < s.
\end{equation}
Then
\begin{align}
    \nonumber \rho_{0}(s;A)
    &= \frac{1}{s} \int \frac{\dd[2]{k_\parallel}} {(2\pi)^2} \sum_{n\in\mathbb{Z}}
    \int_A^{+\infty} \frac{\dd{a}}{C_w a^4}
    \\ \nonumber
    & \qquad \int_{\theta \in SO(3)} \dd{\mu(\theta)} \omega_{k}a^3 \abs{\Tilde{w}(a\mat{R}^{-1}(\theta)\bm{k})}^2 
    \\
    &= \frac{1}{s} \int \frac{\dd[2]{k_\parallel}} {(2\pi)^2} \sum_{n\in\mathbb{Z}} \omega_k \Tilde{f}(A\abs{\bm{k}}) \label{eq:VaccumEnergyNoEuMac}
\end{align}
where
\begin{equation}
    \Tilde{f}(A\abs{\bm{k}}) = \int_A^{+\infty} \frac{\dd{a}}{C_w a} \int_{\theta \in SO(3)} \dd{\mu(\theta)} \abs{\Tilde{w}(a\mat{R}^{-1}(\theta)\bm{k})}^2 
\end{equation}
is the cutoff function $\Tilde{f}$ as in \cref{eq:CutoffFunction} and serves to attenuate high momenta.
This reproduces the standard regularization procedure.
Then one can follow the original derivation by Casimir \cite{casimir1948a} and use the Euler-Maclaurin formula

\begin{align}
    \nonumber\rho_0(s;A)
    &=\frac{2}{s}\left(\frac{F(0;A)}{2}+\sum_{n=1}^{+\infty}F(n;A)\right) \\
    &= \frac{2}{s}\left(\int_{0}^{+\infty}F(n;A)\dd{n} - \sum_{m=1}^{+\infty} \frac{B_{2m}}{(2m)!}F^{(2m-1)}(0;A)\right), \label{eq:NoReEnergy}
\end{align}
where
\begin{equation}
    F(n;A) = \int \frac{\dd[2]{k_\parallel}}{(2\pi)^2}\sqrt{k_\parallel^2 + \left(\tfrac{2n\pi}{s}\right)^2}\Tilde{f}\left(A\sqrt{k_\parallel^2 + \left(\tfrac{2n\pi}{s}\right)^2}\right)\label{eq:AuxFunctionF}
\end{equation}
and $B_n$ is the $n$th Bernoulli number.
The first term in \cref{eq:NoReEnergy} is equal to the energy density of a free field without boundary conditions:
\begin{equation}
    \frac{2}{s}\int_{0}^{+\infty}F(n;A)\dd{n}=\int \frac{\dd[3]{\bm{k}}}{(2\pi)^3}\abs{\bm{k}}\;\Tilde{f}(A\abs{\bm{k}}).
    \label{eq:BulkEnergy}
\end{equation}
This leads to the definition of the renormalized energy density, which can be interpreted as the difference between the energy inside and outside of the plates:
\begin{equation}
    \rho(s;A) = \rho_0(s;A) - \int \frac{\dd[3]{\bm{k}}}{(2\pi)^3} \abs{\bm{k}}\;
    \Tilde{f}(A\abs{\bm{k}}).
    \label{eq:RenormalizedVaccumEnergyNoEuMac}
\end{equation}
Finally, substituting $\rho_0(s;A)$ from \cref{eq:NoReEnergy} obtains
\begin{align}
    \nonumber\rho(s;A) 
    &= - \frac{2}{s}\sum_{m=1}^{+\infty} \frac{B_{2m}}{(2m)!}F^{(2m-1)}(0;A)
    \\ \nonumber
    &= -\frac{\pi^2}{45 s^4} + \frac{8\pi^2}{s^4} \sum_{m=2}^{+\infty}
    \frac{B_{2m+2}}{(2m+2)!} 2m(2m-1) 
    \\
    & \qquad \times \Tilde{f}^{(2m-2)}(0) \left(\frac{2\pi A}{s}\right)^{2m-2}
    \label{eq:PBCEnergyPlusRemainder}
\end{align}
which reduces to the standard result in the limit $A \to 0$, as expected.

\subsection{Dirichlet boundary condition}
In the case of the Dirichlet boundary condition
\begin{equation}
    \phi(t,x,y,0) = \phi(t,x,y,s) = 0,
\end{equation}
the mode structure is slightly different compared to the periodic boundary condition.
In particular, in the expression for the Hamiltonian \cref{eq:WaveletHamiltonian} one has to do the replacements
\begin{equation}
    k_z \to \frac{\pi n}{s},\quad \int \frac{\dd[3]{k}}{(2\pi)^3}\to\sum_{n=1}^{+\infty}\int\frac{\dd[2]{k}_\parallel}{(2\pi)^2}.
\end{equation}
Given this, the regularised vacuum energy density then reads
\begin{align}
    \nonumber\rho_0(s;A) &= \frac{1}{s}\sum_{n=1}^{+\infty} F(n/2;A) \\
    \nonumber&= \frac{1}{s}\bigg( \int_{0}^{+\infty}F(n;A)\dd{n} - \frac{1}{2}F(0;A) \\
    &\quad - \sum_{m=1}^{+\infty} \frac{B_{2m}}{(2m)!2^{2m-1}} F^{(2m-1)}(0;A) \bigg),
\end{align}
where the term $-\frac{1}{2}F(0;A)$ amounts to a constant shift of the total energy due to the boundary conditions \cite{asorey2013} and in this case does not contribute to the Casimir force.
The renormalized vacuum energy density is then
\begin{align}
    \nonumber\rho(s;A) 
    &= - \frac{1}{s}\sum_{m=1}^{+\infty} \frac{B_{2m}}{(2m)!2^{2m-1}} F^{(2m-1)}(0;A)
    \\ \nonumber
    &= -\frac{\pi^2}{720 s^4} + \frac{\pi^2}{2s^4} \sum_{m=2}^{+\infty}
    \frac{B_{2m+2}}{(2m+2)!} 2m(2m-1)
    \\
    & \qquad \times \Tilde{f}^{(2m-2)}(0) \left(\frac{\pi A}{s}\right)^{2m-2},
    \label{eq:DBCEnergyPlusRemainder}
\end{align}
which can also be obtained from the periodic result in \cref{eq:PBCEnergyPlusRemainder} via the replacement $s \to 2s$.

%--------------------------------------------------%

%--------------------------------------------------%
\section{Detecting a scale cutoff}\label{sec:Detection}
The Casimir effect manifests itself via the Casimir force
\begin{equation}
    F_{\mathcal{C}}(s) = -\dv{s}\left(s\rho(s;0)\right).
\end{equation}
To detect a scale cutoff, we consider $\rho(s;A)$ instead of $\rho(s;0)$ and examine the behaviour of the force.
For sufficiently large separation compared to the cutoff, the lowest order term (in $A/s$) from \cref{eq:PBCEnergyPlusRemainder} or \cref{eq:DBCEnergyPlusRemainder} gives a sufficiently precise description.
The series expansion for the force is then
\begin{align}
    F_{\mathcal{C}}^{PB}(s;A) &= -\frac{\pi^2}{15s^4} + \frac{8\pi^2}{s^4}\sum_{m=2}^{+\infty}a_n\left(\frac{2\pi A}{s}\right)^{2m-2}, \label{eq:PBC_Force_Plus_Remainder}\\
    F_{\mathcal{C}}^{DB}(s;A) &= -\frac{\pi^2}{240s^4} + \frac{\pi^2}{2s^4}\sum_{m=2}^{+\infty}a_n\left(\frac{\pi A}{s}\right)^{2m-2},
\end{align}
where
\begin{equation}
    a_n = \frac{B_{2m+2}}{2m+2}\frac{\Tilde{f}^{(2m-2)}(0)}{(2m-2)!}.
\end{equation}
For simplicity, the following discussion is going to focus on the Casimir force in the case of periodic boundary conditions and the scale cutoff being nonzero.
The lowest order correction from \cref{eq:PBC_Force_Plus_Remainder} then reads
\begin{align}
    F_{\mathcal{C}}(s;A) &= -\frac{\pi^2}{15s^4} + \frac{\pi^2}{63s^4}\Tilde{f}^{(2)}(0)\left(\frac{2\pi A}{s}\right)^2 + O(A^4).
\end{align}
Another possibility is to examine the behaviour of the force when the separation of the plates approaches the value of the scale cutoff.
As discussed below, the Euler-Maclaurin formula is not the preferred tool for this.
Both of these regimes strongly depend on the choice of the wavelet.
\subsection{Hermitian wavelets}

\begin{figure}
        \includegraphics[width=.9\columnwidth]{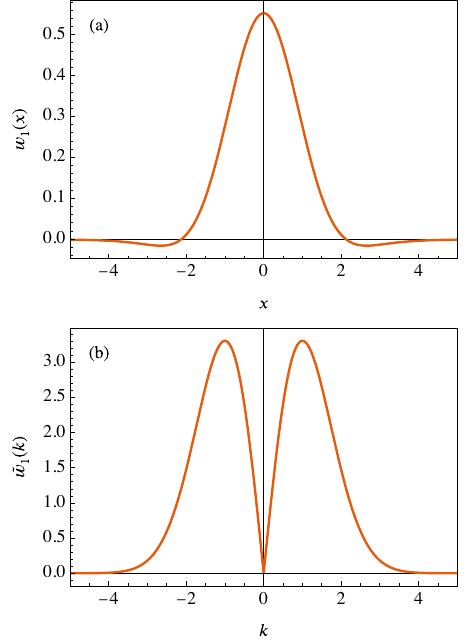}
    \caption{Cross section of the first Hermitian wavelet \cref{eq:HermiteWavelet} in both position (a) and momentum (b) representation.}
    \label{fig:HermiteWaveletCrossSection}
\end{figure}

\begin{figure}
    \centering
    \includegraphics[width=.95\columnwidth]{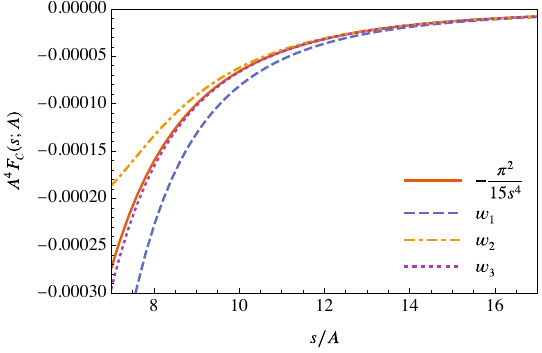}
    \caption{Dependence of the Casimir force $F_\mathcal{C}$ on the separation of the plates $s$.
    The scale cutoff is set $A=1$.
    The solid line corresponds to the usual result $-\frac{\pi^2}{15s^4}$ and the other lines represent the Casimir force perceived by the Hermitian wavelets for different choices of $n$.}
    \label{fig:HermiteWaveletComparison}
\end{figure}

In one dimension, the Hermitian wavelets are usually defined as derivatives of the Gaussian.
Their Fourier image is taken as the definition in higher dimensions so that they automatically satisfy the admissibility condition \cref{eq:AdmissibilityCondition}
\begin{align}
    w_n(\bm{x}) &= \frac{2^\frac{n-1}{2}\Gamma(\frac{3+n}{2})}{\sqrt{\pi\Gamma(\frac{3}{2}+n)}}~_1F_1\left(\frac{3+n}{2};\frac{3}{2};-\frac{\bm{x}^2}{2}\right),\label{eq:HermiteWavelet}
    \\
    \Tilde{w}_n(\bm{k}) &= \frac{2\pi}{\sqrt{{\Gamma(3/2+n)}}}(-i\abs{\bm{k}})^n e^{-\frac{\bm{k}^2}{2}},
\end{align}
here $_1F_1(a;b;z)$ is the Kummer confluent hypergeometric function. The cutoff function \cref{eq:CutoffFunction} for the Hermitian wavelets is
\begin{equation}
    \Tilde{f}_n(k) = \frac{\Gamma(n,k^2)}{(n-1)!} = e^{-k^2}\sum_{l=0}^{n-1}\frac{k^{2l}}{l!}.
\end{equation}
The lowest order correction then reads
\begin{equation}
    F_{\mathcal{C}}(s;A) = -\frac{\pi^2}{15s^4} - \frac{2\pi^2}{63s^4}\delta_{1,n}\left(\frac{2\pi A}{s}\right)^2 + O(A^4).\label{eq:CFHerm}
\end{equation}
Interestingly, for the wavelet with label $n$ the first $n-1$ corrections vanish.
Equivalently, choosing a higher $n$ for the wavelet gives a stronger suppression of the effects caused by the presence of a scale cutoff when the separation is large compared to the cutoff, see \cref{fig:HermiteWaveletComparison}.

\subsection{Wavelet associated with exponential cutoff}
In the usual treatment of the Casimir effect, the exponential regulator is often introduced
\begin{equation}
    \Tilde{f}(k) = e^{-k}.
\end{equation}
The wavelet corresponding to this cutoff function can be found and reads
\begin{align}
    w(\bm{x}) &= 
    \sqrt{\frac{3}{2\pi}}
    \frac{
        \sin\left(\frac{5}{2}\arctan(2\abs{\bm{x}})\right)
    }{
        \abs{\bm{x}}(1+4\bm{x}^{2})^{5/4}
    },\label{eq:ExpCutoffWavelet}
    \\
    \Tilde{w}(\bm{k}) &=
    \frac{\pi}{\sqrt{3}}\sqrt{\abs{\bm{k}}}e^{-\abs{\bm{k}}}.
\end{align}

\begin{figure}
        \includegraphics[width=.9\columnwidth]{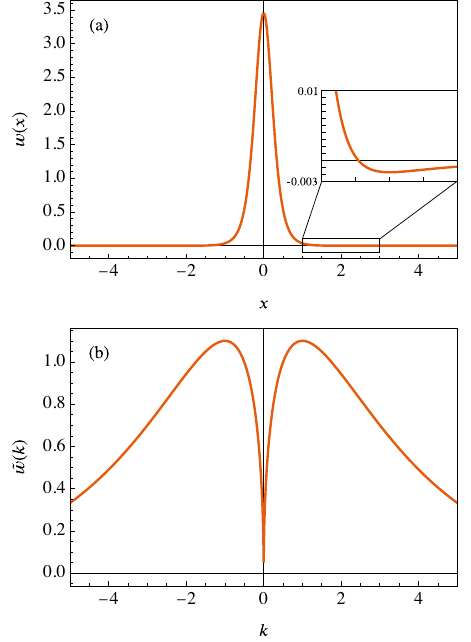}
    \caption{Cross section of the wavelet associated with exponential cutoff \cref{eq:ExpCutoffWavelet} in both position (a) and momentum (b) representation.}
    \label{fig:ExpWaveletCrossSection}
\end{figure}

For this choice of wavelet the lowest order correction in the asymptotic behaviour of large separation compared to the scale cutoff is
\begin{equation}
    F_{\mathcal{C}}(s;A) = -\frac{\pi^2}{15s^4} + \frac{\pi^2}{63s^4}\left(\frac{2\pi A}{s}\right)^2 + O(A^4),\label{eq:CFExp}
\end{equation}
which has an opposite sign compared to the expression for the first Hermitian wavelet \cref{eq:CFHerm}.
Interestingly, in this case, the expression \cref{eq:VaccumEnergyNoEuMac} can be evaluated in a closed form
\begin{equation}
    \rho_0(s;A)=\frac{\coth \left(\frac{\pi  A}{s}\right)}{\pi A^3}+\frac{\pi  A \coth \left(\frac{\pi  A}{s}\right)+s}{A^2 s^3\sinh^2\left(\frac{\pi  A}{s}\right)}.
\end{equation}
The integral for bulk energy \cref{eq:BulkEnergy} evaluates to $\frac{3}{\pi ^2 A^4}$ and thus the renormalized energy reads
\begin{equation}
    \rho(s;A) = \frac{\pi A \coth \left(\frac{\pi  A}{s}\right) - 3}{\pi^2 A^4}+\frac{\pi  A \coth \left(\frac{\pi  A}{s}\right)+s}{A^2 s^3\sinh^2\left(\frac{\pi  A}{s}\right)},
\end{equation}
which in turn gives the exact expression for the Casimir force
\begin{equation}
    F_\mathcal{C}(s;A)=\frac{3}{\pi ^2 A^4}-\frac{\pi ^2 \left(\cosh \left(\frac{2 \pi  A}{s}\right)+2\right)}{s^4\sinh^4\left(\frac{\pi  A}{s}\right)}.\label{eq:CasimirForceExact}
\end{equation}

\begin{figure}
\centering
    \includegraphics[width=.9\columnwidth]{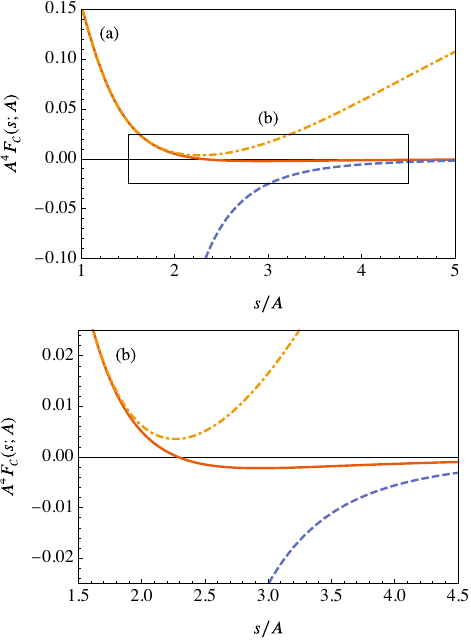}
    \caption{Dependence of Casimir force on the separation of the two plates when the wavelet \cref{eq:ExpCutoffWavelet} is considered.
    The scale cutoff is set $A=1$.
    The solid line corresponds to the exact expression \cref{eq:CasimirForceExact}, the dashed-dotted line on the left corresponds to \cref{eq:RenormalizedVaccumEnergyNoEuMac} where the sum runs between $-3$ and $3$.
    Finally, the dashed line on the right is the expression \cref{eq:CFExp} obtained from the Euler-Maclaurin formula.}
    \label{fig:Orig_vs_Euler_Mac}
\end{figure}

Compare here the application of the Euler-Maclaurin formula versus direct calculation of \cref{eq:VaccumEnergyNoEuMac}.
The Euler-Maclaurin formula is suitable when $A/s$ is small.
In the usual derivation of the Casimir effect, only the dominant term of \cref{eq:PBCEnergyPlusRemainder} is considered and the rest is ignored (essentially taking the $A\to 0$ limit), which recovers the behaviour on length scales far away from the scale cutoff.
When examining length scales comparable to the cutoff, the expansion \cref{eq:PBCEnergyPlusRemainder} starts to lose precision as the sum is usually asymptotic.
Here, the original sum \cref{eq:VaccumEnergyNoEuMac} is preferable to analyse the behaviour around the cutoff.
This is illustrated in \cref{fig:Orig_vs_Euler_Mac}.

\Cref{fig:Orig_vs_Euler_Mac} shows another interesting phenomenon: as the separation approaches the scale cutoff, the force becomes repulsive, indicating reluctance of the plates being closer than the scale cutoff.
Although particular numeric values differ, the same behaviour is also obtained for the Hermitian wavelet family.
The  difference here is that the sum \cref{eq:VaccumEnergyNoEuMac} does not have a closed form expression.

\begin{figure}
\centering
    \includegraphics[width=.95\columnwidth]{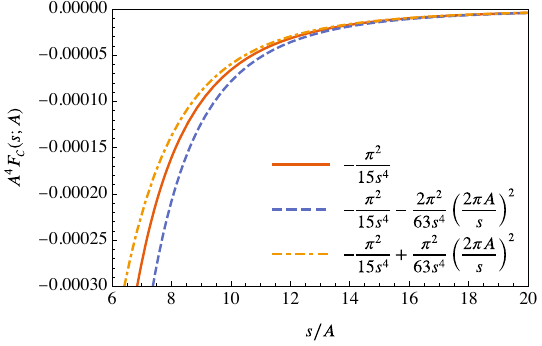}
    \caption{Dependence of the Casimir force on the separation of the two plates with the scale cutoff set to $A=1$.
    The solid line corresponds to the usual result $-\frac{\pi^2}{15 s^4}$, the dashed line is the correction \cref{eq:CFHerm} when the first Hermitian wavelet is considered, and the dotted dashed line is the correction \cref{eq:CFExp} obtained from the wavelet \cref{eq:ExpCutoffWavelet}.}
    \label{fig:ExpansionComparison}
\end{figure}

Thus the scale cutoff is in theory detectable before reaching the critical regime discussed above.
When the separation of the plates is one order of magnitude above the cutoff, the expansions in \cref{eq:CFHerm} and \cref{eq:CFExp} differ meaningfully from the usual value $-\frac{\pi^2}{15 s^4}$.
This is an alternative interpretation of the result in \cite{altaisky2011}. \Cref{fig:ExpansionComparison} illustrates this and also shows that the resulting force depends on the choice of the wavelet.

\subsection{Wavelets with extreme properties}
Here we consider two examples of wavelets whose cutoff function satisfies $\Tilde{f}^{(m)}(0) = 0$ for $m > 0$. 
Such a function could be the following smoothed step function
\begin{equation}
    \Tilde{f}(k) = \left\{\begin{array}{cl}
       1 & 0\leq k \leq 1 \\
       1 - \left(1+\exp\left(\frac{1}{k-1}-\frac{1}{2-k}\right)\right)^{-1}  &  1<k<2\\
       0  & 2 \leq k
    \end{array}\right.,
\end{equation}
where the Fourier image of the corresponding wavelet could be set as
\begin{equation}
    \Tilde{w}_\text{bump}(\bm{k})
    = \mathcal{N} \frac{ \sqrt{\abs{\bm{k}}}\sqrt{2\bm{k}^2-6\abs{\bm{k}} + 5}
        }{
        (\abs{\bm{k}}-1)(2-\abs{\bm{k}}) \cosh\left(\frac{3/2-\abs{\bm{k}}}{(\abs{\bm{k}}-1)(2-\abs{\bm{k}})}\right)
        }
    \label{eq:SmoothedCutoffWavelet}
\end{equation}
in region $1<\abs{\bm{k}}<2$ and zero outside.
The Fourier transform of this wavelet does not appear to have a closed form expression, however the wavelet has several interesting properties.
Firstly, dyadic scaling generates an orthonormal set
\begin{equation} \int_{\mathbb{R}^3}\dd[3]{\bm{x}^\prime} w\left(\frac{\bm{x}^\prime-\bm{x}}{2^k}\right)w\left(\frac{\bm{x}^\prime-\bm{y}}{2^l}\right) = \delta_{kl}.
\end{equation}
However, integer shifts at the same scale or dyadic powers of integer shifts across scales do not produce an orthonormal set and, therefore, cannot be used to generate a basis for a discrete wavelet transform.
There are some shifts that create a function orthogonal to the original, but they do not appear to obey a regular pattern.
The second observation is that the wavelet image of the derivative operator
\begin{equation}
    D_i(a,a',\bm{x},\bm{y}) = \int_{\mathbb{R}^3}\dd[3]{\bm{x}'} w\left(\frac{\bm{x}'-\bm{x}}{a}\right)\partial_i w\left(\frac{\bm{x}' - \bm{y}}{a'}\right)
\end{equation}
has a faster-than-polynomial fall-off in $\abs{\bm{x}-\bm{y}}$.
This means that this wavelet has a much better localisation than, for example, sinc wavelets, while still maintaining orthogonality at dyadic scaling.
It also suggests that it may be useful in a scale-limited analysis of QFT where it is desirable to maintain a local representation of derivative operators (where by ``local'' we mean that the operator representation decays at least superpolynomially).

\begin{figure}
    \centering
    \includegraphics[width=.9\columnwidth]{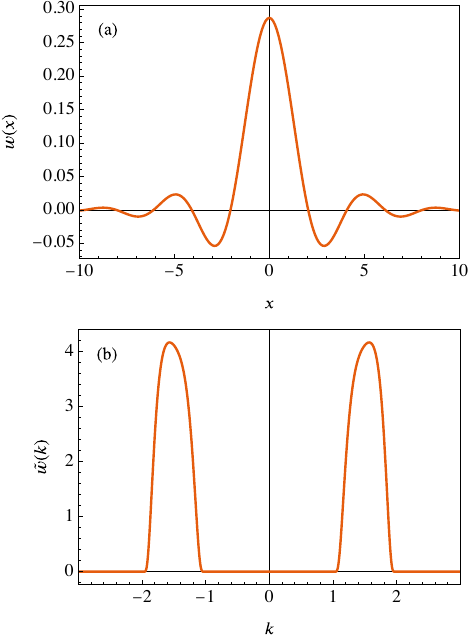}
    \caption{Cross section of the wavelet that is defined as a bump function \cref{eq:SmoothedCutoffWavelet} in the momentum representation (b).
    The cross section in position representation (a) is computed numerically directly from the inverse Fourier integral.}
    \label{fig:bumpwavelet}
\end{figure}

Another possible choice for the cutoff function is a relative of the prototypical example of a non-analytic smooth function:
\begin{equation}
    \Tilde{f}(k) = \left\{\begin{array}{cr}
        1-e^{-\frac{1}{k^{10}}} & k\neq 0 \\
        1 & k=0
    \end{array}\right.,
\end{equation}
where the power $10$ is chosen so that the wavelet is an $L^2$ function in the three-dimensional space.
The expression for the wavelet is then
\begin{align}
    w_\text{Meijer}(\bm{x}) 
    &= \frac{1}{2^{7/10} \abs{\bm{x}} \sqrt{\pi\Gamma(7/10)}} G^{6,0}_{0,11}
    \left( \left. \begin{array}{c} \\ \Vec{b} \end{array} \right|
    \frac{\bm{x}^{10}}{2\cdot 10^{10}} \right)
    \label{eq:NonAnalyticCutoffWavelet}
    \\
    \Tilde{w}_\text{Meijer}(\bm{k}) 
    &= \sqrt{\frac{20\pi^2}{\Gamma(7/10)}} \frac{1}{\abs{\bm{k}}^5}
    \exp\left(-\frac{1}{\bm{k}^{10}}\right),
\end{align}
where $G$ is the Meijer G-function and $\Vec{b} = (\frac{1}{10},\frac{3}{10},\frac{3}{10},\frac{1}{2},\frac{7}{10},\frac{9}{10},0,\frac{1}{5},\frac{2}{5},\frac{3}{5},\frac{4}{5})$.

\begin{figure}
    \centering
    \includegraphics[width=.9\columnwidth]{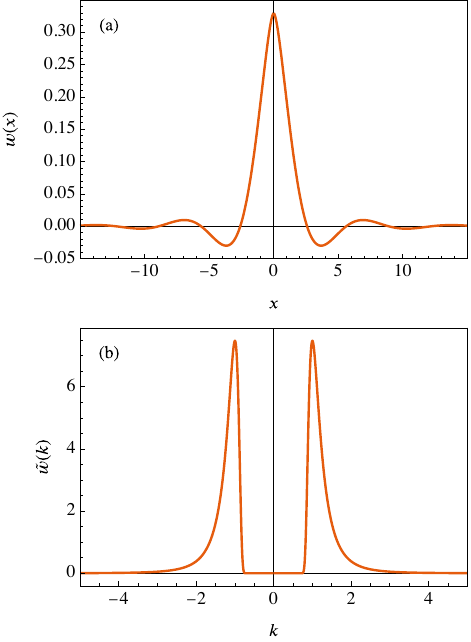}
    \caption{The cross section of the wavelet \cref{eq:NonAnalyticCutoffWavelet} associated with the non-analytic cutoff function $1-e^{-1/k^{10}}$ both in position (a) and momentum (b) representation.
    The momentum representation is zero only for $k=0$, in the neighbourhood of zero the value of the function is so small that the plotting software rounds it down to zero.}
    \label{fig:Non_Analytic_Wavelet}
\end{figure}

The Euler-Maclaurin formula \cite{graham1994} for a $p$ times differentiable function is
\begin{equation}
    \frac{F(0)}{2}+\sum_{n=1}^{+\infty}F(n) = \int_0^{+\infty}F(x)\dd{x} + \sum_{k=1}^{p-1}\frac{B_{k+1}}{(k+1)!}F^{(k)}(0) + R_p,
\end{equation}
where the remainder $R_p$ is given by the formula
\begin{equation}
    R_p=(-1)^{p+1}\int_0^{+\infty}F^{(p)}(x)\frac{B_p(x-\lfloor x\rfloor)}{p!}\dd{x}.
\end{equation}
where $B_n(x)$ is the Bernoulli polynomial of order $n$.
Here, the Euler-Maclaurin formula must be used with care, because the remainder does not go to zero as the order of the derivative $p$ is taken to infinity. The remainder satisfies the difference equation
\begin{equation}
    R_{p+1}-R_p = \frac{B_{p+1}}{(p+1)!}F^{(p)}(0),
\end{equation}
which means that if there exists $p_0$ such that for all $p>p_0$ the derivative $F^{(p)}(0)$ vanishes, the sequence of remainders becomes a constant sequence $R_p = R_{p_{0}}$, for $p>p_0$.
This is precisely the case for the wavelet with a non-analytic cutoff function defined in \cref{eq:NonAnalyticCutoffWavelet}.
The renormalized energy for these wavelets therefore reads
\begin{equation}
    \rho(s;A) = -\frac{\pi^2}{45s^4}+\frac{2}{s}R_4(s;A),
\end{equation}
where the remainder has the form
\begin{equation}
    R_4(s;A) = \frac{(2\pi)^2}{s^3}\int_0^{+\infty}\dv[3]{x}\left(x^2\Tilde{f}\left(\frac{2\pi A x}{s}\right)\right)\frac{B_4(x-\lfloor x\rfloor)}{4!}\dd{x}.
\end{equation}
The Casimir force is then obtained straightforwardly as
\begin{equation}
    F_\mathcal{C}(s;A) = -\dv{s}(s\rho(s;A)) = -\frac{\pi^2}{15s^4} - 2\pdv{s} R_4(s;A).
\end{equation}

\begin{figure}
    \centering
    \includegraphics[width=.95\columnwidth]{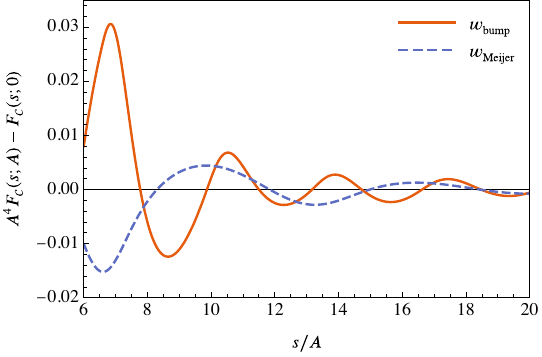}
    \caption{Comparison of the corrections to Casimir force obtained from the wavelet $w_\text{bump}$~\cref{eq:NonAnalyticCutoffWavelet} associated with cutoff function $\Tilde{f}(k)=1-e^{-1/k^{10}}$ (dashed) and the wavelet $w_\text{Meijer}$~\cref{eq:SmoothedCutoffWavelet} associated with the smoothed cutoff function (solid), both exhibiting oscillatory behaviour.
    The scale cutoff is set to $A=1$.}
    \label{fig:Osc_Casimir}
\end{figure}

These wavelets therefore have the interesting property that the asymptotic behaviour is isolated from the infinite sum and all of the short-range behaviour is contained in the remainder term.
The resulting Casimir force is oscillatory, as shown in \cref{fig:Osc_Casimir}, which ought to be particularly useful in the context of detecting the existence of a scale cutoff.
This oscillatory behaviour can also be obtained with higher-order Hermitian wavelets.

%--------------------------------------------------%
\section{Discussion and Conclusion}

We demonstrated in \Cref{sec:CasimirEffect} the use of wavelet regularization to introduce a scale cutoff to a canonical field theory calculation, and showed using several known and some new  wavelet families that the cutoff manifests itself as higher-order terms that disappear in the zero-scale limit.
This serves to give further credence to the usefulness of wavelet-based regularization in QFT.

In \Cref{sec:Detection} we demonstrated the relationship between the choise of wavelet and the character of the resulting Casimir force.
This suggests the potential to 'tune' a measuring probe to either enhance or minimise the effects of a scale cutoff in a system.
If the objective is to minimise the effects of a scale cutoff (i.e. to more closely approximate the continuum limit using a finite scale) then this can be achieved by engineering the probe so that the aperture function corresponds to a higher order Hermitian wavelet (see \cref{fig:HermiteWaveletComparison}).
In contrast, if the intention is to maximise the deviation of the measurements of the scale-limited force from the continuum force, so as to reveal the presence of the scale cutoff, then a wavelet with a non-analytic cutoff function would provide this.
In such a system the Casimir force would then exhibit distinctive oscillatory behaviour as a function of separation.

Our analysis also provides a new perspective for modelling fundamental cutoffs.
The GUPs mentioned in the introduction model a minimal length scale by modifying the commutation relations between the position and momentum operators \cite{kempf1995,kempf1997}.
The electromagnetic field operators are then expanded in terms of maximally localised states, which regularises the theory and one then obtains an attractive correction \cite{frassino2012} to the Casimir force.
Our approach is distinct in that, because the maximally localised states fail to satisfy the wavelet admissibility condition, they cannot be used to generate a wavelet frame.
We show that via wavelet-based regularization one encounters positive, negative, and oscillatory corrections to the force, however, we have observed in all our examples that as the separation approaches the scale cutoff the force becomes repulsive.
This observation may be one possible avenue for further investigation.

\begin{acknowledgements}
We gratefully acknowledge discussions with Nicholas Funai, Achim Kempf, Dominic Lewis, and Nicolas Menicucci.
We acknowledge the Wallamattagal people of the Dharug nation, whose cultures and customs have nurtured, and continue to nurture, the land on which some of this work was undertaken: Macquarie University.
We acknowledge support from the ARC Centre of Excellence for Engineered Quantum Systems (EQUS, CE170100009).
D. J. G. and G. K. B. acknowledge support from the Australian Research Council (ARC) through Grant No. DP200102152.
S. V. and D. J. G. were supported by the Sydney Quantum Academy, Sydney, Australia.
\end{acknowledgements}

\bibliography{main}

\end{document}